# MISSION STATEMENTS IN UNIVERSITIES: READABILITY AND PERFORMANCE

- Julián David Cortés-Sánchez: Distinguished Professor – School of Management, Universidad del Rosario, Bogotá, Colombia. Visiting Scholar – Fudan University, China.
- Katerina Bohle Carbonell: Postdoctoral Fellow - Whitaker Institute, National University Ireland. PhD, Organizational Behavior and Education - Maastricht University, Netherlands
- Liliana Rivera: Associate Professor – School of Management, Universidad del Rosario, Colombia. PhD, Logistics and Supply Chain Management - Massachusetts Institute of Technology.

# MISSION STATEMENTS IN UNIVERSITIES: READABILITY AND PERFORMANCE

**Abstract**

The mission statement(s) (MS) is one of the most-used tools for planning and management. Universities worldwide have implemented MS in their knowledge planning and management processes since the 1980s. Research studies have extensively explored the content and readability of MS and its effect on performance in firms, but their effect on public or nonprofit institutions such as universities has not been scrutinized with the same intensity. This study used Gunning's Fog Index score to determine the readability of a sample of worldwide universities' MS and two rankings, i.e., Quacquarelli Symonds World University Ranking and SCImago Institutions Rankings, to determine their effect on performance. No significant readability differences were identified in regions, size, focus, research type, age band, or status. Logistic regression (cumulative link model) results showed that variables, such as universities' age, focus, and size, have more-significant explanatory power on performance than MS readability.

## 1. Introduction

The implementation of mission statement(s) (MS) as a planning and management tool in the higher education sector dates to the 1980s (Kotler and Murphy, 1981). Organizations use MS to communicate their higher aims and goals to their internal and external stakeholders (Campbell, 1989). For a MS to comprehensively express its guidelines, it needs to consider four components: i.e., purpose (why the organization exists); values (what the organization believes in); standards and behaviors (the policies and behavioral patterns that guide the organization); and strategy (the organization's strategy for achieving its goals) (Campbell, 1989). Pearce (1982) proposed eight additional key components to enhance the comprehensiveness of a MS: i.e., targeted customers; basic products or services; primary markets; principal technologies; concerns for survival, growth, and profitability; company philosophy; company self-concept; and concern for its public image. The MS must be clearly communicated to achieve a positive relationship between the spirit of the MS and the organization's resulting performance.

This line of inquiry has produced a considerable amount of research in the past 30 years. A meta-analysis of more than 20 years of evidence supported the significant, although small, positive association between MS and financial performance (Desmidt et al., 2011). It remains unclear whether this finding also holds for nonprofit organizations. Research on universities' MS have analyzed the impact of their contents (Ellis and Miller, 2014) and effect on the universities' identity and their employees' and students' behavior and ethical orientation (Atkinson, 2008; Hladchenko, 2013). However, MS must be clearly understood by the reader to have an impact on them. A broader understanding of the effect of MS communication factors such as readability on performance remains understudied. Also, universities are considered integral actors in both the knowledge economy (Goddard and Chatterton, 1999) and national innovation systems (Etzkowitz and Leydesdorff, 2000; Freeman, 1995); thus, innovation efficiency evaluations on performance-related outputs (e.g., research papers and patents) generated by universities are critical to the strategic growth of both the higher education sector (Altbach and Knight, 2007; Cabeza-Pullés et al., 2018; Glass et al., 2006; Schofer and Meyer, 2005) and the national productive ecosystem (Franco & Haase, 2015; Grigg, 1994; Plewa et al., 2013).

The research question that guides this study is: does the readability of universities' MS influence their performance? This study aims to determine the effect of readability on performance in a sample of worldwide universities. The rest of the paper is organized as follows: Section 2 presents a literature review of MS and performance. Section 3 presents the methodology, including the universities' baseline search list (i.e., Quacquarelli Symonds World University Ranking [QS-WUR])

Gunning's Fog Index (GFI) score calculation (Gunning, 1969), and performance indicators based on the QS-WUR and the SCImago Institutions Rankings (S-IR). Section 4 explores if there are significant differences in readability by regions, size, focus, research type, age band, and status in universities using one-way analysis of variance (ANOVA), and present the logistic regression results to determine the effects of readability on universities' performance. Section 5 presents the discussion, which is followed by conclusions and recommendations in Section 6.

**2. Literature review**

One of the first studies aimed to examine the relationship between MS and organizational performance was conducted by Pearce and David (1987). They found that firms with a comprehensive MS showed a higher performance. Subsequent evidence showed the positive effect of MS contents, readability, and planning processes on employees' satisfaction and behavior (Baetz and Bart, 1996; Bart and Baetz, 1998), performance on both financial and mission achievement measures in firms (Bart and Hupfer, 2004; Cortés-Sánchez and Rivera, 2019; Godoy-Bejarano and Tellez-Falla, 2017), and hospitals' achievement of goals (Bart and Tabone, 1999). One of the few comprehensive studies conducted for manufacturing SMEs in developing countries (Turkey) (Duygulu et al., 2016) found that MS's components such as survival, growth, and profit; philosophy and values; and public image, were significantly associated with production performance auto-reporting scores (product returns, and losses).

In contrast, studies also have found no significant difference between the return on assets for firms with and without MS (Bart and Baetz, 1998), no correlation between MS and performance in percentage growth in sales volume (O'Gorman and Doran, 1999), and the added value per employee was the only performance indicator associated with the absence of a MS g(Valerij, 2012). In their meta-analysis summarizing 20 years of research on the topic, Desmidt et al. (2011) supported the positive association, albeit small, between MS and financial performance, considering both hard (e.g., profit margin, 10-year average earnings per share) and soft (i.e., managers' perception of employee satisfaction) financial indicators. However, the reported relationship depends on operational decisions, the performance measures were only financially related, and most studies were conducted in private organizations. It remains unclear whether the same association between MS and performance exists for other types of organizations (e.g., nonprofit or public), and additional insights considering knowledge-intensive indicators (e.g., research papers or patents) instead of exclusively financial indicators.

Compared with private organizations, universities pursue a societal mission (Collini, 2012): i.e., educating society and creating knowledge. However, universities have come under increasing pressure to provide evidence of their societal benefits (Cuthill et al., 2014). Given universities' societal importance and the potential impact a well-crafted MS can have on performance, investigating universities' MS leads to a better understanding of their performance drivers. Studies related to universities' MS aimed to understand their content (Ellis and Miller, 2014; Firmin and Gilson, 2010; Kosmützky, 2012; Kosmützky and Krücken, 2015; Seeber et al., 2017), the effect on their identity and employees' behavior (Atkinson, 2008; Hladchenko, 2013), response capacity to political, cultural, and economic factors (Deus et al., 2016; Hladchenko, 2016; Kuenssberg, 2011), and differences among private or public institutions (Efe and Ozer, 2015; Morphew and Hartley, 2006). However, additional studies analyzing the effect of the content or readability of MS on strategic activities, such as corporate communication and performance, remain scarce.

To the best of our knowledge, Cochran and David (1986) conducted the first research study considering MS inspiring content and readability and their effect on corporate communications in Fortune 500 companies and business schools from the United States (US) accredited by the AACSB (Association to Advance Collegiate Schools of Business). The study used the same readability index used here (i.e., Gunning Fog Index). Two independent raters assessed the MS inspiring tone. The study found that only 50% of the MS were inspiring; companies' MS used a positive tone over business schools; and both evidenced lower readability with at least 24.6 years of education needed in average. At the bottom line, both companies and business schools should improve their MS readability. A subsequent study showed differences in performance (e.g., operating budget per full-time faculty, percentage of faculty with a doctorate) related to the MS content in US business schools (Short and Palmer, 2008). Concretely, findings regarding the impact of MS readability on performance had presented mixed results based on the different geographic locations of the investigated organizations and differences in performance metrics.

The above review of the relationship between MS content and readability and performance in firms and universities identified four gaps. The first is the absence of transnational studies, while the second is a disengagement from the global South because most studies were conducted in the US. The third is the narrow scope of universities' performance measurements beyond operating budgets per full-time faculty, percentage of faculty with a doctorate, and indexed publications. Finally, the fourth gap is the limited attention on MS readability as a crucially active component for both internal and external stakeholders to understand and interiorize the organization's purpose and course of action, and as an input for employees' day-to-day actions (Sattari et al., 2011).

## 3. Methodology

### 3.1. Data

#### 3.1.1. Mission statements

The QS-WUR was used to identify an available list of universities' MS to consider for analysis because it is one of the most discussed university ranking systems in the recent literature (Anowar et al., 2015; Hou and Jacob, 2017; Huang, 2012; Moed, 2017; Soh, 2015, 2017). Universities with no score on academic and employer's reputation were excluded, i.e., both factors add up to 50% of the overall score (QS-WUR, 2019). Two research assistants searched for each university's MS on their institutional website. The MS gathered were explicitly described as "mission" or "mission statement." MS with titles such as "values," "purpose," or "vision" were not considered due to a more-precise data gathering method following the data collection steps outlined by Desmidt et al. (2011). Only MS in English were considered due to language standardization. Consequently, the sample was reduced from 400 to 248 MS. Table 1 presents the number of universities and MS by continent and country, while Table 2 presents the number and percentages of the universities by size, focus, research, age band, and status.

**Table 1 Number of universities by continent and number of MS by country**

| Key | (Sub)Continent | Universities | Key | Country | MS |
|---|---|---|---|---|---|
| EU | Europe | 94 (38%) | UK | United Kingdom | 32 |
| | | | DE | Germany | 14 |
| | | | NL | Netherlands | 8 |
| | | | FI | Finland | 7 |
| | | | CH | Switzerland | 5 |
| | | | BE | Belgium | 5 |
| | | | IE | Ireland | 4 |
| | | | DK | Denmark | 3 |
| | | | ES | Spain | 3 |
| | | | PT | Portugal | 3 |
| | | | FR | France | 2 |
| | | | NO | Norway | 2 |
| | | | SE | Sweden | 1 |
| | | | IT | Italia | 1 |
| | | | AT | Austria | 1 |
| | | | EE | Estonia | 1 |
| | | | PL | Poland | 1 |
| | | | GR | Greece | 1 |
| NA | North America | 79 (31%) | US | United States | 69 |
| | | | CA | Canada | 10 |
| AS | Asia | 48 (19%) | JP | Japan | 11 |
| | | | CN | China | 6 |
| | | | HK | Hong Kong | 5 |
| | | | IN | India | 5 |
| | | | KR | Korea | 5 |
| | | | MY | Malaysia | 5 |
| | | | TW | Taiwan | 3 |
| | | | SA | Saudi Arabia | 2 |
| | | | SG | Singapore | 2 |
| | | | TH | Thailand | 1 |
| | | | ID | Indonesia | 1 |
| | | | KZ | Kazakhstan | 1 |
| | | | LB | Lebanon | 1 |
| OC | Oceania | 18 (7%) | AU | Australia | 13 |
| | | | NZ | New Zealand | 5 |
| LA | Latin America | 5 (2%) | CO | Colombia | 2 |
| | | | BR | Brazil | 1 |
| | | | CL | Chile | 1 |
| | | | AR | Argentina | 1 |
| AF | Africa | 4 (1%) | ZA | South Africa | 3 |
| | | | EG | Egypt | 1 |
| | | | | Total MS | 248 |

*Source:* The author, based on QS-WUR (2013) and university websites. This table presents the sample of universities by continent and countries.

**Table 2 Number and percentage of universities by size, focus, research, age band, and status.**

| Metric | Key | Meaning | #Us | % |
|---|---|---|---|---|
| Size | S | Small | 6 | 2% |
| | M | Medium | 47 | 19% |
| | L | Large | 140 | 56% |
| | XL | Extra-large | 55 | 22% |
| Focus | FC | Fully comprehensive | 160 | 65% |
| | CO | Comprehensive | 71 | 29% |
| | FO | Focused | 14 | 6% |
| | SP | Specialist | 3 | 1% |
| Research | MD | Medium | 4 | 2% |
| | HI | High | 30 | 12% |
| | VH | Very high | 214 | 86% |
| Age band | H | Historic (>100 years) | 162 | 65% |
| | MA | Mature (50–100 years) | 53 | 21% |
| | E | Established (25–50 years) | 24 | 10% |
| | Y | Young (10–24 years) | 9 | 4% |
| Status | A | Public | 207 | 83% |
| | B | Private | 41 | 17% |

*Source:* The author, based on QS-WUR (2013). This table presents the sample of universities by size, focus, research output, age band, and status.

### 3.1.2. Readability indices

Readability indices are used to determine the comprehension difficulty of written material (Flesch, 1948) and avoid needless complexities in the mechanics of writing (Gunning, 1969). The overall finding is that good readability is related to better performance, clear disclosure, and improved understanding (Linsley and Lawrence, 2007; Rennekamp, 2012). The most-used index for readability in the management and finance English-language research literature is the Gunning's Fog Index (GFI) (Clark et al., 1990; Flory et al., 1992; Kaminski and Clark, 1987; Karlinsky and Koch, 1983; Lo et al., 2017; Loughran and McDonald, 2014). The GFI estimates the number of years of schooling a person needs to understand a given text on the first reading (e.g., 17 equals to a college graduate). The GFI is calculated as follows:

$$GFI = 0.4(ASL + PHW)$$

Where average sentence length (ASL) is the number of words and syllables in a text of at least 100 words, and the total number of words is divided by the total number of sentences. Percent of hard words (PHW) is the number of words that contain more than three syllables (nonproper noun combinations of easy or hyphenated words, or two-syllable verbs made into three-syllable verbs by adding -es and -ed endings and divided by the total number of words in the text).

### 3.1.3. Performance

Two rankings to measure universities' performance were considered: QS-WUR and S-IR. The QS-WUR was selected because it is one of the most-discussed rankings in grecent literature. QS-WUR evaluates universities based on six metrics: (1) academic reputation; (2) employer reputation; (3) faculty/student ratio; (4) citations per faculty; (5) international faculty ratio; and (6) international student ratio. According to Shehatta and Mahmood (2016) the QS-WUR and five major university rankings, such as the Academic Ranking of World Universities, the Times Higher Education World University Ranking, the US News & World Report Best Global University Rankings, the National Taiwan University Ranking, and the University Ranking by Academic Performance, showed moderate-to-high internal correlations among them regardless of their methodological differences. Other than QS-WUR and other well-known university ranking systems, S-IR has not been discussed comparatively with QS-WUR. S-IR assesses more than 5,000 institutions in terms of research (i.e., output, international collaboration, normalized impact, high-quality publications, excellence, scientific leadership, excellence with leadership, scientific talent pool), innovation (i.e., innovative knowledge, technological impact), and societal impact (i.e., web sites and inbound links) (Bornmann et al., 2012; Jeremić et al., 2013). The detailed definitions and input for each S-IR indicator are presented in Annex 1.

The periods considered for both rankings were 2013 to 2015 for QS-WUR and 2013 to 2017 for S-IR. Several years were considered for testing how consistent the effect of MS readability is on the organization's performance. The S-IR publishes the ranking position for each examined institution online (1 to 5,200). Therefore, inverse min–max normalization from 0 to 100, the former as the highest score (closer to one), was conducted to create a common scale for both rankings. A permanent link to the databases is available at http://bit.ly/2XWMALl or by using a QR code (Fig. 1).

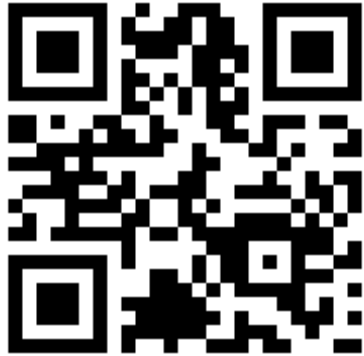

**Fig. 1. QR code for dataset access.**

**4. Results**

**4.1. Readability**

The average GFI was 20.82 (SD = 11.38), ranging from 1.2 to 142.8. A GFI of 17 or higher indicates that a reader would require a graduate college education to understand clearly the MS at first reading; therefore, the reader needs to have completed college education to be able to understand the average university's MS.

Let us consider the extreme and mean values of the GFI. First, the MS of *Technische Universität Wien* (Germany), "*Technology for the people*," has a GFI of 1.2, which means that a person below sixth grade in the US education system, would understand it at a first read. Second, the MS of the Arizona State University (USA) has a GFI of 142.8. That means that the reader would need a graduate college education to understand it at first read. At 326 words, this MS is very long and it is also one of the few to include several quantitative goals (e.g., "enhance research competitiveness to more than $815 million in annual research expenditures"). Finally, an example of an average MS would be that of the Indian Institute of Technology, Delhi (India) with a GFI of 20.8:

> To generate new knowledge by engaging in cutting-edge research and to promote academic growth by offering state-of-the-art undergraduate, postgraduate and doctoral programs. To identify, based on an informed perception of Indian, regional and global needs, areas of specialization upon which the institute can concentrate. To undertake collaborative projects which offer opportunities for long-term interaction with academia and industry. To develop human potential to its fullest extent so that intellectually capable and imaginatively gifted leaders can emerge in a range of professions.

We began by investigating if specific university characteristics present significant differences on their GFI. One-way analyses of variance (ANOVA) were performed to determine significant differences in GFI means by regions, size, focus, research type, age band, and status in universities. The categories that included too few universities were excluded from the analysis. There were no significant differences detected. First, regarding regions, universities were divided into four groups: Europe, North America, Asia, and Oceania. There was no statistically significant difference at the $p < 0.05$ level in GFI scores [$F(3, 235) = 0.96$, $p = 0.41$]. Second, regarding size, universities were divided into three groups: medium, large, and extra-large. There was no statistically significant difference at the $p < 0.05$ level in GFI scores [$F(2, 238) = 0.67$, $p = 0.51$]. Third, regarding focus, universities were divided into three groups: i.e., fully comprehensive, comprehensive, and focused. There was no statistically significant difference at the $p < 0.05$ level in GFI scores [$F(2, 240) = 0.20$, $p = 0.81$]. Fourth, regarding research type, universities were divided into two groups: very high and high. There was no statistically significant difference at the $p < 0.05$ level in GFI scores [$F(1, 242) = 0.01$, $p = 0.94$]. Fifth, regarding age band, universities were divided into three groups: historic (>100 years old), mature (50–100 years old), established (between 25–50 years old). There was no statistically significant difference at the $p < 0.05$ level in GFI scores [$F(2, 236) = 0.01$, $p = 1.00$]. Sixth, regarding status, universities were divided into two groups: private and public. There was no statistically significant difference at the $p < 0.05$ level in GFI scores [$F(1, 246) = 0.57$, $p = 0.45$].

## 4.2. Readability and performance

Table 3 represents the regression results for GFI as an independent variable and universities' overall performance scores as dependent variables, controlling for university characteristics. First, the analyses were conducted with the QS-WUR (QS_osc) and subsequently with the S-IR (Sci_rnk). In both cases, our dependent variable was the university ranking, with lower values indicating a better position in the ranking. The independent variable was readability as operationalized through the GFI scores.

As our dependent variable is a ranked variable, we conducted a logistic regression analysis using a cumulative link model. To run this model, we first grouped the ranking into four groups: top performers (rank 1 to 50), high performers (rank 50 to 100), average performers (rank 100 to 200), and low performers (rank 200 and higher). Subsequently, we collapsed certain descriptors because our data did not contain enough sample points for each performance year and category combination. Specifically, we collapsed the small and medium-size category into one, resulting in three categories (small to medium, large, and extra-large). We combined the categories of focused, comprehensive, and specialist into one category: i.e., not fully comprehensive. Considering age bands, we combined the three youngest age categories, mature, established, and young, to create two age bands: i.e., established and historic. Finally, we excluded the research category because of too-few data points in two out of three categories.

In the logistic regression analysis, the test for proportional odds was only valid for the focus and age categories. Therefore, we ran a partial proportional odds model using size and status categories as nominal variables. The results (Table 3) indicated that the QS-WUR ranking in 2013 was predicted by GFI ($\beta = -0.07$, $p = 0.01$) and age group ($\beta = -1.01$, $p = 0.01$). However, the effect of GFI on performance is small. Nevertheless, this indicates that universities with lower GFI, i.e., having MS that are easier to understand, achieved lower rankings. That is, for universities of the same size and status, the greater readability of their MS decreased their ranking position by 7%. Regarding university descriptors, the odds indicate that historic universities have a 64% chance to achieve a better ranking position compared with established universities. Hence, universities whose MS had a higher GFI (lower readability) and who existed for a long time had better ranking positions than a university whose MS has a lower GFI (higher readability) and was founded recently.

While this might seem counterintuitive, it could indicate that universities are targeting a specific audience in their MS and consider that complexity, as expressed through a higher GFI, resonates with their target audience. Older universities would have a more established presence and have more resources that enable them to achieve a higher-ranking position. Older universities benefit from a first-mover advantage (FMA) and had several years to build up their reputations. However, in the subsequent year, GFI did not relate to the ranking position. The ranking was related to age ($\beta = -0.74$, $p = 0.01$ in 2014; $\beta = -0.70$, $p = 0.02$ in 2015) and a fully comprehensive focus of university ($\beta = -1.29$, $p = 0.00$ in 2014; $\beta = -1.14$, $p = 0.00$ in 2015). Potentially, the university's focus provides sufficient guidance to employees that the readability of the MS does not add substantial explanatory power.

**Table 3 Results of logistic regression predicting QS-WUR ranking position.**

|  |  | QS-WUR ranking (2013) |  | QS-WUR ranking (2014) |  | QS-WUR ranking (2015) |  |
|---|---|---|---|---|---|---|---|
|  |  | $\beta$ (St. E) | Odds | $\beta$ (St. E) | Odds | $\beta$ (St. E) | Odds |
| GFI |  | −0.07 (0.03)* | 0.93 | −0.01 (0.01) | 1.01 | −0.01 (0.01) | 0.99 |
| Focus: fully comprehensive |  | −0.33 (0.48) | 0.72 | −0.74 (0.3)* | 1.63 | −0.70 (0.29)* | 0.48 |
| Age: historic |  | −1.01 (0.4)* | 0.36 | −1.29 (0.29)*** | 1.01 | −1.14 (0.28)*** | 0.28 |
|  |  |  |  |  |  |  |  |
| Threshold coefficients |  |  |  |  |  |  |  |
| 1–50\|50–100 | Intercept | −4.24 (0.84) |  | 0.5 (−7.42) |  | −3.46 (0.48) |  |
| 50–100\|100–200 | Intercept | −3.02 (0.77) |  | 0.45 (−5.67) |  | −2.62 (0.44) |  |
| 100–200\|200–384 | Intercept | −1.12 (0.72) |  | 0.42 (−2.87) |  | −1.14 (0.41) |  |
| 1–50\|50–100 | Intercept | 3.01 (1.14) |  | 0.49 (0.98) |  | 0.07 (0.48) |  |
| 50–100\|100–200 | Status: public | −0.42 (0.68) |  | 0.41 (−0.47) |  | −0.16 (0.41) |  |
| 100–200\|200–384 | Status: public | −0.98 (0.65) |  | 0.4 (-0.38) |  | −0.32 (0.38) |  |
| 1–50\|50–100 | Size: medium | −1.93 (1.25) |  | 0.54 (0.87) |  | 0.89 (0.51) |  |
| 50–100\|100–200 | Size: medium | 0.64 (0.63) |  | 0.44 (0.94) |  | 0.44 (0.44) |  |

| | | | | |
|---|---|---|---|---|
| 100–200\|200–384 | Size: medium | −1.06 (0.62) | 0.4 (−0.29) | −0.06 (0.38) |
| 1–50\|50–100 | Size: extra-large | 0.69 (0.56) | 0.45 (1.06) | 0.04 (0.46) |
| 50–100\|100–200 | Size: extra-large | 0.59 (0.47) | 0.36 (0.42) | 0.53 (0.36) |
| 100–200\|200–384 | Size: extra-large | 0.68 (0.6) | 0.36 (0.36) | 0.04 (0.35) |
| | | | | |
| N | | 121 | 236 | 248 |
| AIC | | 306.38 | 591.11 | 608.75 |

*Note.* QS-WUR ranking = Quacquarelli Symonds World University Ranking; GFI = Gunning's Fog Index. Categories: focus (fully comprehensive, not comprehensive), age (mature, established), status (public, private), size (medium, large, extra-large). *** $p < 0.01$, ** $p < 0.01$, * $p < 0.05$. This table presents the results of the logistic regression analysis for GFI as an independent variable and universities' overall performance scores as dependent variables for QS-WUR.

When running the same test for the other performance indicator, S-IR, we transformed the performance variable into the same groups as in the previous analysis. The proportional odds assumptions were met for all years and all categorical variables. The results are reported in Table 4. The readability of the MS did not predict the ranking in any time period. However, certain university descriptors had a significant impact on performance. Extra-large size and age consistently predicted ranking position for each year between 2013 and 2017 (size: $\beta = -1.4$, $p = 0.00$ in 2013, $\beta = -1.27$, $p = 0.00$ in 2014, $\beta = -1.27$, $p = 0.00$ in 2015, $b = -1.32$, $p = 0.00$ in 2016, $\beta = -1.45$, $p = 0.00$ in 2017; age: $\beta = -1.24$, $p = 0.00$ in 2013, $\beta = -1.29$, $p = 0.00$ in 2014, $\beta = -1.32$, $p = 0.00$ in 2015, $\beta = -1.44$, $p = 0.00$ in 2016, $\beta = -1.2$, $p = 0.00$ in 2017). The odds that the ranking position of established universities compared with mature universities was higher are between 76% and 60%. Extra-large universities had a 73% to 77% chance of achieving a better ranking due to their size. In all years except for 2013, a fully comprehensive focus had a positive relationship with performance ($\beta = -0.69$, $p = 0.03$ in 2014; $\beta = -0.81$, $p = 0.01$ in 2015; $\beta = -1$, $p = 0.00$ in 2016; $\beta = -0.92$, $p = 0.01$ in 2017), increasing the odds of a better ranking position between 50% and 63%. Universities with a public status achieved a better ranking position in 2013 and 2017 ($\beta = -0.77$, $p = 0.04$ and $\beta = -0.82$, $p = 0.03$, respectively).

**Table 4**

Results of logistic regression analyses predicting S-IR position.

|  | S-IR (2013) |  | S-IR (2014) |  | S-IR (2015) |  |
|---|---|---|---|---|---|---|
|  | $\beta$ (St. E) | Odds | $\beta$ (St. E) | Odds | $\beta$ (St. E) | Odds |
| GFI | −0.01 (0.01) | 0.99 | −0.01 (0.01) | 0.99 | −0.01 (0.01) | 0.99 |
| Focus: fully comprehensive | −0.61 (0.31) | 0.54 | −0.69 (0.32)* | 0.50 | −0.81 (0.32)* | 0.44 |
| Age: established | −1.42 (0.32)*** | 0.24 | −1.29 (0.31)*** | 0.28 | −1.32 (0.32)*** | 0.27 |
| Status: public | −0.77 (0.38)* | 0.46 | −0.65 (0.37) | 0.52 | −0.67 (0.38) | 0.51 |
| Size: medium | 0.36 (0.4) | 1.43 | 0.35 (0.4) | 1.43 | 0.44 (0.41) | 1.55 |
| Size: extra-large | −1.4 (0.33)*** | 0.25 | −1.27 (0.32)*** | 0.28 | −1.27 (0.32)*** | 0.28 |
| N | 240 |  | 240 |  | 240 |  |
| AIC | 533.37 |  | 539.32 |  | 530.39 |  |

|  | S-IR (2016) |  | S-IR (2017) |  |
|---|---|---|---|---|
|  | $\beta$ (St. E) | Odds | $\beta$ (St. E) | Odds |
| GFI | −0.01 (0.01) | 0.99 | −0.01 (0.01) | 0.99 |
| Focus: fully comprehensive | −1 (0.33)** | 0.37 | −0.92 (0.33)** | 0.40 |
| Age: established | −1.44 (0.33)*** | 0.24 | −1.26 (0.33)*** | 0.28 |
| Status: public | −0.66 (0.38) | 0.51 | −0.82 (0.38)* | 0.44 |
| Size: medium | 0.56 (0.43) | 1.75 | 0.51 (0.43) | 1.67 |
| Size: extra-large | −1.32 (0.32) *** | 0.27 | −1.45 (0.33)*** | 0.23 |
| N | 239 |  | 238 |  |
| AIC | 514.58 |  | 509.10 |  |

*Note.* AIC, Akaike's information criterion; S-IR, SCImago Institutions Ranking; GFI, Gunning's Fog Index. Categories: focus (fully comprehensive), age (mature, established), status (public, private), size (medium, large, extra-large). *** $p < 0.01$, ** $p < 0.01$, * $p < 0.05$. This table presents the results of the logistic regression analysis for GFI as an independent variable and universities' overall performance scores as dependent variables for S-IR.

## 5. Discussion

The results show that universities with lower MS readability had a better position in the QS-WUR ranking in 2013, which could be explained from two perspectives. First, the goal of MS is to align organizational processes to achieve a common goal. A comprehensive MS provides a sense of direction and purpose to employees and helps their managers to align their decision-making (Campbell, 1989). Sattari et al. (2012) observed that poor readability of MS represents a challenge for effective communication among the firms' employees. Thus, readability becomes an issue when there is a need to communicate with an audience having a variety of education levels effectively (Clark et al., 1990). In this sense, Sattari et al. (2012) reported that the MS of Fortune 500 companies require the reading level of university graduates, which excludes a significant portion of the population from understanding these MS. The same seems to hold for universities. The human capital of universities has a high level of educational attainment. For example, around 43% and 57% of doctoral graduates in the UK and the EU, respectively, are employed by higher education institutions (HEIs) (CRAC, 2013). Essentially, the majority of university employees would not have any difficulty in reading and understanding MS that require an education level higher than high school.

Second, university MS may serve other purposes other than guiding performance. Morphew and Hartley (2006) observed that a university's MS serves as a communication mechanism for stakeholders, instead of addressing internal performance. Furthermore, the clients (stakeholders) of universities differ from those of regular companies. For instance, Cortés-Sánchez (2018) performed a transnational content analysis of universities' MS and found an overall emphasis on society and students as stakeholders. Fârte (2013) noted that HEI customers, such as students, differ from typical business customers because these stakeholders will become future sources of change in organizations worldwide. To justify its existence, each university, faculty, or department should focus on their target audiences, among other things, i.e., their principal stakeholders. An ideal university MS should consider employees, academics, administrators, society, students, other research institutions, other universities, government, and graduates (Yilmaz, 2012). As mentioned earlier, Pearce (1982) claims that an integral MS should involve eight components: i.e., the product or service; the primary market; the technologies used; the company's goals, philosophy, and self-concept; the company's public image; and the stakeholders. Therefore, it is necessary to use technical, clear, and inspirational language to motivate not only employees, but also stakeholders to pursue the university's strategic goals (Bugaj and Rybkowski, 2018). Consequently, universities'

MS must be more comprehensive, vigorous, and probably longer and more complex than those of other organizations.

In comparing the results between the two university rankings used in this study, our results show that readability proved to be a more significant predictor of QS-WUR instead of S-IR rankings, which may be explained by their differences in societal (stakeholder) orientation. Indeed, a 50% QS-WUR ranking accounts for both employers' satisfaction and academic reputation. In contrast, S-IR's societal component (web visibility) accounts for only 20%, while research and innovation accounts for 70%.

The results also suggest a strong relationship among universities' performance and features, such as age, size, and focus. Concerning age, we found that historical universities had better performance in educational rankings. Established universities who entry and position first in the market hold FMA (Li et al., 2011), which emerges mainly from market share considering its effects on performance compared with their competitors (Short and Payne, 2008; Suarez and Lanzolla, 2007; VanderWerf and Mahon, 1997). The results of FMA lead to privileged access to valuable production spaces and resources, as well as the prompt generation of knowledge, practices, and loyalty (Carpenter and Nakamoto, 1989). Universities with FMA also experience proprietary learning effects, patents, preemption of input factors and location, and the development of buyer switching costs (Lieberman and Montgomery, 1987). Additionally, Bowman and Bastedo (2011) affirm that rankings give great importance to reputation and prestige, which corresponds at some level to an anchoring effect. Consequently, universities with a long history have a greater chance of enjoying advantages related to their reputation based on the amount of time they had to gain recognition and reputation and forge a strong community of graduates (Morgeson and Nahrgang, 2008).

As for size, extra-large universities perform better in rankings. In this sense, Li et al. (2011) express the possibility of economies of scale on larger universities' activities, which might imply growth in their relative output and enable higher performance. Jump (2014) observed that governments also have an interest in merging and investing in HEIs because greater size usually leads to greater visibility as well as greater efficiency. These public resource holders usually push for infrastructure developments, such as the number of laboratories, researchers, databases, and resources, which enable better university performance. This observation is also in line with Cortés-Sánchez (2018), who claimed that MS from extra-large-, large-, and medium-sized universities gave priority to the term "research" over "education" in response to their aim to become research-based universities instead of knowledge-based universities. However, Hazelkorn (2009) claims that global rankings usually favor the larger and older institutions with a broader scope on the offered subjects (comprehensive focus), which allows them to accumulate physical, financial, intangible, and human capabilities. As for focus, we found that comprehensive universities perform better than those with a narrower approach. In that line, Crespi (2007) noted that universities active in many scientific disciplines have higher research productivity due to the existence of economies of scope in research, which might influence their performance.

## 6. Conclusions

University MS require comprehensiveness and a wider range of text to communicate their purpose and course of action to their stakeholders. More specifically, MS that are more complex in terms of their employed words and concepts, and are hence more difficult to read, represent a more comprehensive and complete description of the university's purpose. That is a suitable description for internal and external stakeholders and helps delimit the organizational purpose and guide the decision-making process. The high educational attainment required for this exercise can be found in the context of universities, in either faculty staff, administrators, students, or governmental institutions in charge of assessing and monitoring higher education systems.

The complex readability of MS was related to increasing performance, although in the particular context of QS-WUR rankings. This exceptional relationship could be explained by the 50% composition of the QS-WUR overall score consisting of two factors: i.e., employers' satisfaction and academic reputation. Therefore, universities develop a comprehensive MS to state their purpose and course of action to employers in the private, public, and third sectors, and their academic peers. By extension, this increases their reputation and acknowledgment score performance. While such a MS could help communicate the university's purpose and attract students and staff, the reported relationship between MS readability and performance was only significant in 2013. That raises the question about what other contextual factors, such as political and economic climate, could help explain this relationship.

Variables such as antiquity, focus, and size had a consistent and significant explanatory power for university performance in both university ranking systems after controlling for diverse university characteristics. In the case of historical universities, they hold a FMA that allows them to access a privileged place in the market. Their consistent performance over time would allow those institutions to gain a better reputation among their stakeholders. Considering extra-large universities, higher performance could be unsurprisingly related to economies of scale, public funding (i.e., laboratories, researchers, grants), and research focus, which are variables with relevant weights in both university ranking systems. A similar remark can be stated regarding comprehensive universities because the diversity of disciplines and departments has the potential to attract more students from either national or international origins and researchers with varied know-how, which generates good perceptions from different stakeholders in both the private and public sectors with research output generated in several fields.

A limitation of this study is that university performance was measured through university rankings. The limitations arise through methodological weaknesses in how these rankings are computed. Concerns have been raised about the ranking construction, criteria, assumptions, and lack of visibility. Others have pointed out that the rankings might have a field bias due to the questionnaires and surveys that focus on a small number of fields and favor universities that focus on those areas and encourage universities to broaden their range of subjects to increase their performance. The findings of this study could be contrasted by investigating other performance criteria, such as drop-out rates, student employability, private-public funding, and university–corporate research collaborations. Such performance measures are very much context-dependent and would require the inclusion of macro-level information, such as national expenditure research and development activities, national institution quality, or the sophistication of the private sector, among other economic and institutional environment factors.

**Annex 1**

The complete information presented in this Annex was obtained from the SCImago Institution Ranking methodology website: https://www.scimagoir.com/methodology.php

Research factor:

- Normalized Impact (NI) is computed over the institution's leadership output using the methodology established by the Karolinska Institutet in Sweden where it is named "Item oriented field normalized citation score average." The normalization of the citation values is done on an individual article level. The values (in decimal numbers) show the relationship between an institution's average scientific impact and the world average set to a score of 1. Size-independent indicator.

- Excellence with Leadership (EwL) indicates the amount of documents in excellence in which the institution is the main contributor.

- Output (O) refers to the total number of documents published in scholarly journals indexed in Scopus. Size-dependent indicator.

- International Collaboration (IC) is the Institution's output produced in collaboration with foreign institutions. The values are computed by analyzing an institution's output whose affiliations include more than one country address. Size-dependent indicator.

- High-Quality Publications (Q1) is the number of publications that an institution publishes in the most influential scholarly journals of the world. These are those ranked in the first quartile (25%) in their categories as ordered by SCImago Journal Rank (SJRII) indicator. Size-dependent indicator.

- Excellence (Exc): Excellence indicates the amount of an institution's scientific output that is included in the top 10% of the most cited papers in their respective scientific fields. It is a measure of high-quality output of research institutions. Size-dependent indicator.

- Scientific Leadership (L) indicates the amount of an institution's output as main contributor, i.e., the amount of papers in which the corresponding author belongs to the institution. Size-dependent indicator.
- Scientific Talent Pool (STP) is the total number of different authors from an institution in the total publication output of that institution during a particular period of time. Size-dependent indicator.

Innovation factor:

- Innovative Knowledge (IK) is the scientific publication output from an institution cited in patents. Based on PATSTAT (http://www.epo.org). Size-dependent.
- Technological Impact (TI) is the percentage of the scientific publication output cited in patents. This percentage is calculated considering the total output in the areas cited in patents, which are the following: Agricultural and Biological Sciences; Biochemistry, Genetics and Molecular Biology; Chemical Engineering; Chemistry; Computer Science; Earth and Planetary Sciences; Energy; Engineering; Environmental Science; Health Professions; Immunology and Microbiology; Materials Science; Mathematics; Medicine; Multidisciplinary; Neuroscience; Nursing; Pharmacology, Toxicology and Pharmaceutics; Physics and Astronomy; Social Sciences; Veterinary. Based on PATSTAT (http://www.epo.org). Size-independent.
- Patents (PT) are the number of patent applications (simple families). Based on PATSTAT (http://www.epo.org). Size-dependent.

Societal impact factor:

- Number of Backlinks (BN) is the number of networks (subnets) from which inbound links to the institution website came from. Data extracted from ahrefs database (https://ahrefs.com). Size-dependent.
- Web size (WS): number of pages associated to the institution's URL according to Google (https://www.google.com). Size-dependent.